\pdfoutput=1
\documentclass[english, 12pt]{article}
\usepackage{graphicx}
\usepackage{tabularx}
\usepackage{geometry}
\usepackage{babel}
\title{Quantum Compiler Optimizations}
\author{Jeff Booth boothjmx@cs.washington.edu}

\begin{document}

\maketitle

\begin{abstract}
A quantum computer consists of a set of quantum bits upon which operations
called gates are applied to perform computations. In order to perform
quantum algorithms, physicists would like to design arbitrary gates
to apply to quantum bits. However, the physical limitations of the
quantum computing device restrict the set of gates that physicists
are able to apply. Thus, they must compose a sequence of gates from
the permitted gate set, which approximates the gate they wish to apply -
a process called \emph{quantum compiling}.

Austin Fowler proposes a method \cite{Fowler2011} that finds optimal
gate sequences
in exponential time, but which is tractable for common problems.
In this paper, I present several optimizations to this algorithm.
While my optimizations do not improve its overall exponential behavior,
they improve its empirical performance by one to two orders of magnitude.

\end{abstract}

\section{Background}

In classical computing, we can generally rely on the correctness of
hardware because of the size of the circuit components. For example,
if an atom on a hard disk drive changed its spin orientation, or lost
an electron, the hard drive\textquoteright{}s functionality would
not be impaired because it takes many thousands of atoms to represent
and store a single bit of data. However, in quantum computing, data
is stored in quantum bits, which are represented by tiny particles
like trapped ions. These qubits are very easy to perturb, potentially
corrupting calculations based on them. Thus, we use redundancy, in
the form of error-correcting codes, to minimize the impact of individual
errors.

The Steane code is one representation of a quantum bit. It uses seven
physical qubits to represent one Steane code qubit, and can tolerate
an arbitrary error in one of the seven qubits. We can perform any
desired operation on a Steane code qubit by applying a combination
of $H$ (Hadamard) and $T$ gate operations \cite{Nielsen2000}.
$T$ gates are generally complicated to implement in quantum computing hardware,
so we seek to use a minimal number. For practical purposes, in addition to
$H$, we can use the Pauli X operator $X$, the Pauli Z
operator $Z$, the single qubit phase gate $S$, and its inverse $S^{\dagger}$
\cite{Fowler2010}. The gates
$H$, $X$, $Z$, $S$, and $S^{\dagger}$ generate a group under multiplication,
called the Clifford group. Thus, any sequence of gates we choose will
alternate between a member of the Clifford group and a $T$ gate.
A $T^{\dagger}$ gate is also used in this implementation, bringing the total
number of non-identity gates in Fowler's gate set to 25.

A \emph{single-qubit quantum compiler} finds sequences of gates which yield
matrices that are \textquotedblleft{}close\textquotedblright{} to a gate we
would like to apply to a quantum bit. Each gate has a corresponding
matrix that represents the operation it would perform on a quantum
bit. How close one gate is
to another is given by the \textquotedblleft{}Fowler\textquotedblright{}
distance:

\begin{equation}
  \label{eq:fowler_dist}
  dist(U,U_l)=\sqrt{\frac{2-\left| tr\left(U \cdot U_l^\dagger \right) \right|}{2}}
\end{equation}

The longer the gate sequence is, the more closely it can approximate
a desired target gate that is not in the universal instruction set.
However, a longer gate sequence takes more time to compute
on a real quantum computer, increasing the probability of a computation
error. An optimal quantum compiler will find gate sequences which:

\begin{enumerate}
  \item have a minimal Fowler distance from the target gate.
  \item have a minimal length.
\end{enumerate}

\section{Fowler\textquoteright{}s Algorithm}

Austin Fowler presents an algorithm that iterates over sequences in
order from smallest to largest \cite{Fowler2010}. For each sequence,
it multiplies
the sequence gates' matrices together to generate a $2 \times 2$
unitary matrix representing
the complete operation that sequence would perform. The simple brute-force
iteration runs in time exponential in sequence length, since
all sequences of length $n$ are produced by appending all elements of the
universal instruction set to all sequences of length $n-1$.

To reduce the run time, Fowler\textquoteright{}s algorithm intelligently
skips redundant sequences. The algorithm creates a list of unique
sequences for all sequences of length $N$. Then, for each sequence $S$
of length $N+1$, it searches for sub-sequences of length $N$. If a sub-sequence
$Y$ is not in the list, then it is not unique. That means it performs
the same operation as a sequence $V$ which is in the unique sequence
list. Since $Y$ and $V$ are the same, then if you were to replace $Y$ with
$V$ in your sequence $S$, you would get a sequence $W$ that does the same
thing $S$ does.

Since the algorithm iterates over all sequences of length $N$, it will
encounter $W$ anyway (or it has already encountered $W$). Therefore, it
should skip sequence $S$. In fact,
it should increment the sub-sequence $Y$ until it is a unique sequence
$U$. Fowler\textquoteright{}s algorithm contains a tree lookup structure
which, for any given sequence, records the next unique sequence $U$.
The algorithm can determine what sequence to skip to by simply accessing
this tree. It still requires time exponential in sequence length,
but interesting results can now be obtained in mere days using consumer
computer hardware.

As I will demonstrate later, Fowler's tree data structure requires memory
that scales exponentially with the sequence length. Thus, the algorithm
consists of two stages:

\begin{enumerate}
  \item During the first stage, it builds the structure
        until it stores all unique sequences up to length $W$, where $W = 15$
        for most of the experiments.
  \item After the structure is built, it enters the second stage,
        where it generates sequences -- and uses the structure
        to skip them -- but it doesn\textquoteright{}t add them to the structure.
\end{enumerate}

This dramatic change in behavior between stages explains some interesting
features in the following graphs. Also, it means I can only infer behavior
on longer sequences from behavior in the second stage, which explains my focus
on data produced during that stage.

\section{Experimental Goal}

In order to empirically measure the impact of my optimizations, I
need a consistent experimental goal to test on every version of the
algorithm. For this research, I chose to approximate the $\frac{\pi}{6}$
gate $\exp(i\frac{\pi}{12}\sigma_z)$ to $10^{-7}$ accuracy.
Since this approximation is currently very time-consuming, I can use it
to empirically evaluate the impact of my enhancements.

\section{Existing Performance}

In order to better understand the performance characteristics of the
Fowler algorithm, I modified its C source code to obtain performance-related
statistics. In this section, I present the data I gathered, along
with some explanations for unusual data features and speculations
on how the statistics should change for a meaningful performance improvement.
Most of these benchmarks ran on an Amazon Elastic Compute Cloud Medium
computer, which contains a 2-2.4 GHz processor and 3.75 GB of memory.

\subsection{Code Profiling}

I ran a profiler (\texttt{gprof}) to determine where the performance
bottlenecks are. Initially, I thought that
memory accesses would dominate the program\textquoteright{}s runtime,
because of the size of the data structures involved. However, the
program spends 92.49\% of its time inside mathematical functions,
meaning that calculation is the dominant operation. In the first stage
of the algorithm, the program spends 98.5\% of its time checking for
unique matrices. In the second stage, it spends most of its time multiplying
gate matrices together to calculate the matrix for a given sequence.

\subsection{Calculation Time vs. Fowler Distance}

The figure below indicates how much time will be required to obtain
a given Fowler distance using Fowler\textquoteright{}s original source
code. For the purposes of this paper, the Fowler source code is
\textquotedblleft{}unoptimized\textquotedblright{},
as it does not contain my optimizations and is a baseline for comparison.
This graph is perhaps the most important graph of them all, since
we often want gates with a certain specific precision.

\centerline{
\includegraphics[width=.5\textwidth]{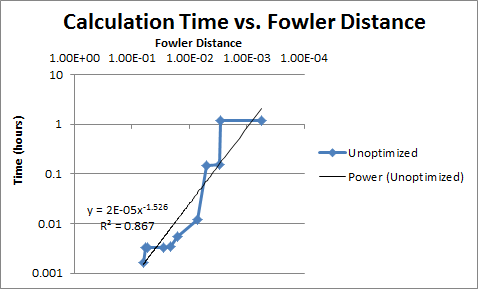}
}

Since there aren\textquoteright{}t very many unique distances, there
are not enough data points to establish a clear trend. A power function
appears to fit the data somewhat closely, though. This power function
predicts that the unoptimized version of the program will take about
110 years to approximate the gate to a distance closer than $10^{-7}$!
This massive exponential expansion explains why Fowler\textquoteright{}s
original paper had no gates with a precision better than $10^{-4}$, since
it would take at least a day for the $\frac{\pi}{6}$ gate to compile to
even that precision!

\subsection{Time vs. Sequence Length}

This metric is related to the above metric because longer sequences
tend to have better precision. However, the relationship between time
and sequence length is much clearer, as can be witnessed by the much
smoother curve. While this graph may not have as much practical significance,
it is much easier to relate this graph to the underlying implementation
of the algorithm.

\centerline{
\includegraphics[width=.4\textwidth]{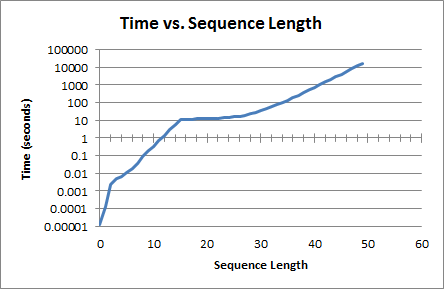}
\includegraphics[width=.4\textwidth]{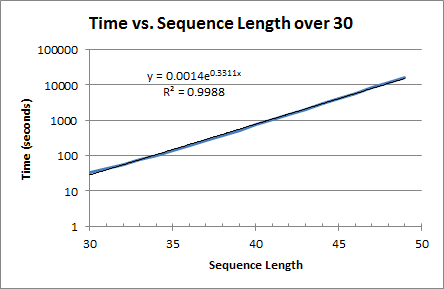}
}

From sequence length 0 to 2, the line has a steep slope. This feature
probably exists because the processor cache has not warmed up yet.
Between 2 and 15, every sequence generated by the algorithm is checked
against a list of unique sequences, to see if it\textquoteright{}s
unique. This check only occurs up to a certain sequence length: 15
in this case. After that, the algorithm speeds up very rapidly until
it reaches about a sequence length of 30. Then, the graph becomes
a clean exponential curve.

To improve performance, I will effectively need to shift this curve
down, producing longer sequences in less time.

\subsection{Unique Sequences Per Sequence Length}

\centerline{
\includegraphics[width=.4\textwidth]{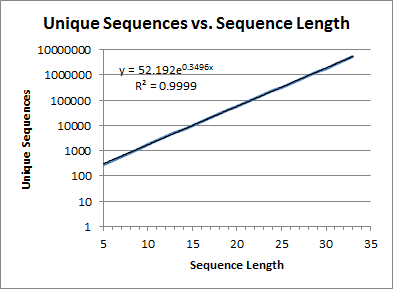}
}

This metric provides insight into the algorithm\textquoteright{}s
storage requirements. It is clear that Fowler\textquoteright{}s optimizations
have not altered the fundamental exponential nature of the problem.
For sequences longer than about 3, the number of unique sequences
grows exponentially with the sequence length. Since I am more worried
about time rather than space, I will not mind if this curve shifts
up. However, I do need to make sure that my optimizations do not consume
too much memory.

\section{Ways to Improve Performance}

To optimize the performance, I need to:
\begin{enumerate}
  \item Speed up calculations such as matrix multiplication.
  \item Reduce the number of calculations required for a given gate
        sequence length.
\end{enumerate}

There are quite a few possible approaches to approaches 1 and 2. Some
of these approaches were taken this quarter, yet others will be left
for future work.

\subsection{\textquotedblleft{}Meet in the Middle\textquotedblright{} Bidirectional Search}

A traditional \textquotedblleft{}uni-directional\textquotedblright{}
search seeks a path from a start state to a goal state by starting
from the start state and exploring all possible paths. A bidirectional
search starts searching from the goal state as well. Thus, the search
paths will \textquotedblleft{}meet in the middle\textquotedblright{}:
each search only has to take $\frac{N}{2}$ steps to meet the other
search. Thus,
instead of taking $O\left(a^N\right)$ time, the algorithm only takes
$O\left(a^{\frac{N}{2}}\right)$ time.
One will need some data structure to store the paths, but inserting
into this data structure does not require exponential time. Thus,
for a given amount of time, the algorithm could compute gate sequences
that are twice as long. This approach is the most promising, and it
was implemented in software.

\subsection{Optimized Unique Matrix Lookup}

The algorithm checks to see if a matrix is unique by calculating the
distance between it and all other matrices. Since 98.5\% of the application\textquoteright{}s
run time is spent in this function, optimizing it could yield significant
improvements in performance in the first stage. However, in the second
stage, no more unique matrix checks are performed; therefore, no time
will be spent in this function. Unless the first stage lasts a long
time, it may not be worth the implementation trouble. This optimization
was easy to implement since the C++ standard template library provides
a red-black binary search tree.

\section{Bidirectional Search}

Searching for the correct gate is like searching through nodes in
a tree: for a given sequence of gates, the computer must choose which
gate to add to the sequence to come closer to the target gate. In
the diagrams below, the arrows represent a choice of gate, and the
boxes represent matrices. When an arrow is drawn from some box A to
a box B, box B is the matrix resulting from multiplying A by some
gate matrix.

\centerline{
\includegraphics[width=.8\textwidth]{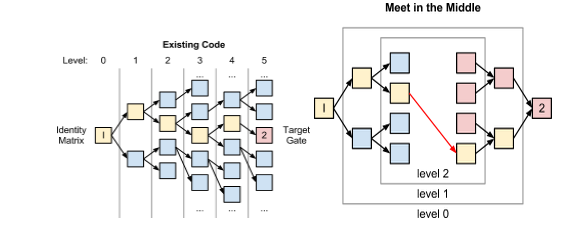}
}

In the example shown in these figures, the existing code must go through
five levels of searching in order to reach the target gate.
At each new level, the algorithm considers adding all of the available gates
to \emph{each} sequence generated by the previous level.
Thus, each step multiplies the number of matrices to consider
by 25. So, for a sequence of length N, there will be $25^N$ operations.
The \textquotedblleft{}meet in the middle\textquotedblright{} figure
reveals that starting the search from the start and the goal results
in the computer exploring fewer levels. Each side would
only have to explore half as many levels since the searches meet in
the middle. Instead of $25^N$ operations, the computer can ideally
perform $2\cdot25^{N/2}$ operations using the MITM (meet in the middle)
algorithm.

\subsection{The Search Index}

The critical component of the MITM algorithm is the structure that
allows the paths to connect. This structure effectively creates the
red arrow in the MITM figure above, matching up left matrices with
right matrices. It must be designed carefully to ensure optimal performance
of the algorithm. For a given left matrix, it should find a minimal
number of right matrices which are close to the left matrix. Thus,
the data structure needs a way to parameterize all of the matrices
stored in it, using parameters that are related to the Fowler distance
between two matrices.

The simplest approach is to choose some reference matrix M, and store
the right matrices in a tree map, using their distances from M as
keys. Then, to find right matrices that are \textquotedblleft{}close\textquotedblright{}
to a left matrix L, the algorithm simply measures the distance from
L to M, and performs a range query for all right matrices that have
about the same distance to M. This trick works because the Fowler
distance measure obeys the triangle inequality: if two matrices L
and R are within some distance d of each other, then the difference
in their distances to some other matrix M will not be greater than
d. In the figure below, this fact is true for all matrices inside
the circle.

\centerline{
\includegraphics[width=.4\textwidth]{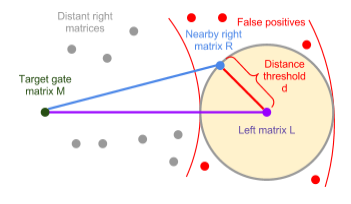}
}

For my implementation, I use the target gate as the reference matrix,
and I choose d to be $10^{-10}$ less than the smallest distance found so
far. Since the left matrix must check its distance from the target
gate anyway, we can re-use the distance calculation without having
to cache it. Note that it is possible for two matrices to be far away
from each other while still having the same distance to M. Thus, the
range query may return false positives, which are shown between the
red lines in the figure. The triangle inequality property simply guarantees
that the range query will not leave out potential candidates.

\subsection{Building the Structure}

For each sequence S the algorithm generates,
a corresponding matrix M is generated. M represents the transformation
that S would perform on a quantum bit. The algorithm usually assumes
that S is a prefix of the solution, meaning that other gates will
be added to the end of S to reach the target gate G. However it\textquoteright{}s
also possible to consider S as a suffix, in which gates are added
onto the beginning of S. In this case, S would work backwards from
G, attempting to come close to the identity matrix, rather than the
other way around. If the computer knows M, it can work backwards by
multiplying the inverse of M with G to get a matrix M2. Then, prefix
sequences can see if S is their suffix by comparing their matrices
to M2. If a prefix matrix is close to M2, then it would be close to
G if it were multiplied by M.

Therefore, the middle structure simply needs to store as many matrices
N as possible, with pointers to their corresponding sequences. It
stores a list of binary search trees by sequence length, so that all
short sequences can be examined before long sequences.

The middle structure only has so much room to store entries, though.
Since the number of unique sequences scales exponentially with the
sequence length, the structure can store entries up to some length
L before running out of memory. Thus, the MITM algorithm does not
always cut the number of search levels in half; instead, it subtracts
L from the number of search levels required to find a solution. This
approach replaces the $O(25^N)$ cost of exploring sequences of length
N with a $O(25^{N-L})$ cost, since a well-optimized middle structure
should not have an exponential lookup time.

\subsection{Performing the Search}

Whenever the algorithm finds a new unique sequence P, it checks the
middle structure to see if one of the suffixes S can connect it to
the target gate G. Since suffixes are searched by ascending length,
the first result should be of optimal length. The search function
is given a distance parameter that indicates the maximum tolerable
Fowler distance for the match; all matrices that are farther away
are skipped. If a result is found, the search function also returns
the distance D from P\textquoteright{}s matrix to S\textquoteright{}s
matrix, so that the distance threshold can be reduced to $D - \epsilon$
(some small value). That way, future searches will only return more
precise matches.

One problem that I noted after obtaining my results is that the real
sequence may not be of optimal length. The Clifford group contains
elements that are composed of multiple real gates, but each Clifford
group element is considered to be one gate in this algorithm. Since
every sequence alternates between Clifford group elements and T gates,
the number of real gates in the sequence of length n returned by the
algorithm is about $n/2 + 3(n/2)$. However, the resulting sequence will
still have an optimal real length: the Clifford group elements are
ordered such that the ones comprised of multiple real gates are visited
later by the algorithm, meaning they are added to the structure at
a later time. Thus, if the structure uses a stable sort, these longer
sequences will be considered later. I am not entirely certain that
my structure does so, however, which would be a good topic for future
research.

Another potential problem is that a very good suffix may be skipped
because a \textquotedblleft{}sufficient\textquotedblright{} suffix
was encountered first. For speed, the MITM algorithm returns the first
suffix that is within the desired distance threshold. Technically,
if this event occurs, the improved suffix would be discovered at the
next search level, so this problem should not impact correctness.
However, that means the best result might not be returned as early
as possible. One sufficient correction would be to continue the search;
it won\textquoteright{}t impact performance because new sequences
are rarely found. This fix could be implemented in future work.

\subsection{Results}

As the graph below shows, the \textquotedblleft{}meet in the middle\textquotedblright{}
(MITM) optimization improved performance by an order of magnitude.
Instead of taking about one hour to calculate a gate sequence that
is within $10^{-3}$ of the target gate, it takes about ten minutes. The
Unoptimized and MITM Width 15 lines both used a \textquotedblleft{}width\textquotedblright{}
of 15, meaning that the middle structure and Fowler\textquoteright{}s
data structures stored sequences of length 15. The actual improvement
appears to depend on the width of the middle structure: when sequences
of length 30 are stored in it, the time is cut by two orders of magnitude
instead of one.  \emph{Note: "unoptimized" refers to Fowler's existing
algorithm without the MITM optimization, not to a simple brute-force
enumeration.}

\centerline{
\includegraphics[width=.6\textwidth]{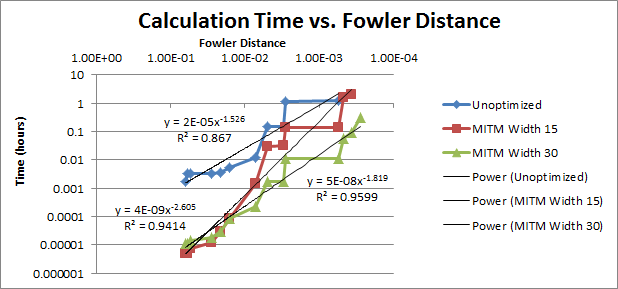}
}

Fowler\textquoteright{}s unoptimized algorithm also improves
performance when the width is increased, because his data structures
can cache more data. Thus, it makes sense that increasing the width
to 30 from 15 results in a larger improvement than just turning on
the MITM optimization.

The memory requirements are much clearer as well: the number of unique
sequences increases exponentially with the sequence length. I omitted
data for sequences of length less than five because they adversely
affect the exponential curve fit.

\centerline{
\includegraphics[width=.6\textwidth]{unique_sequences_vs_sequence_lengths.png}
}

Finally, I noticed that the number of sequences per unit of time was
much larger in the optimized versions than in the unoptimized versions,
confirming my hypothesis. It clearly makes sense to keep expanding
the middle structure if possible: beyond sequences of length 30, the
MITM implementation with width 15 slows down relative to the implementation
with width 30. However, in the long run, the MITM optimization does
not change the base of the exponential that governs the algorithm
run time: notice that all of the lines are roughly parallel towards
the right side of the graph.

\centerline{
\includegraphics[width=.6\textwidth]{time_vs_sequence_length_new.png}
}

I managed to approximate the $\frac{\pi}{6}$ gate to $6.8\times10^{-5}$ precision in about
3 hours and 5 minutes. The result is 72 gates long:

\centerline{$
\begin{array}{c}
HTHT(HS)THTHTHT(HS)THT(HS)T(HS) \\
T(HS)T(HS)THTHT(HS)THTHT(HXZ) \\
THTHTHTHTHT(HS)THTHT(HS)THT(HS) \\
THTHTHTHT(HS)THT(HXS)T^{\dagger}
\end{array}
$}

\section{Change of Basis}

Since the Fowler distance is phase independent, we can adjust gates
to remove their global phase.
Thus, it is possible to represent a quantum gate in $SU\left(2\right)$ by
using just four real numbers.
In the equation below, $\sigma_x$, $\sigma_y$, and $\sigma_z$ are
the Pauli basis matrices.  Since they are multiplied by $i$, the basis
is called the \emph{modified Pauli basis}.

\begin{eqnarray}
  \label{eq:newbasis}
  A &=&
    a_0 \cdot I +
    a_1 \cdot \sigma_x +
    a_2 \cdot \sigma_y +
    a_3 \cdot \sigma_z
  \\
  &=&
    a_0 \cdot \left(
      \begin{array}{cc}
        1 & 0 \\
        0 & 1
      \end{array}
    \right) +
    a_1 \cdot \left(
      \begin{array}{cc}
        0 & i \\
        i & 0
      \end{array}
    \right) +
    a_2 \cdot \left(
      \begin{array}{cc}
        0  & 1 \\
        -1 & 0
      \end{array}
    \right) +
    a_3 \cdot \left(
      \begin{array}{cc}
        i & 0 \\
        0 & -i
      \end{array}
    \right)
  \\
  &=&
    \left(
      \begin{array}{cc}
         a_0 + a_3 \cdot i     & a_2 + a_1 \cdot i \\
        -a_2 + a_1 \cdot i     & a_0 - a_3 \cdot i
      \end{array}
    \right)
\end{eqnarray}

One advantage of this new basis is that the trace distance between two
gates $A$ and $B$ is just double the dot product of their vectors
$a$ and $b$:

\begin{eqnarray}
  tr \left( A \cdot B^{\dagger} \right) &=&
    tr \left(
      \left(
        \begin{array}{cc}
           a_0 + a_3 \cdot i     & a_2 + a_1 \cdot i \\
          -a_2 + a_1 \cdot i     & a_0 - a_3 \cdot i
        \end{array}
      \right)
      \cdot
      \left(
        \begin{array}{cc}
           b_0 + b_3 \cdot i     & b_2 + b_1 \cdot i \\
          -b_2 + b_1 \cdot i     & b_0 - b_3 \cdot i
        \end{array}
      \right) ^ {\dagger}
    \right)
  \\
  &=&
    tr \left(
      \left(
        \begin{array}{cc}
           a_0 + a_3 \cdot i     & a_2 + a_1 \cdot i \\
          -a_2 + a_1 \cdot i     & a_0 - a_3 \cdot i
        \end{array}
      \right)
      \cdot
      \left(
        \begin{array}{cc}
          b_0 - b_3 \cdot i     & -b_2 - b_1 \cdot i \\
          b_2 - b_1 \cdot i     &  b_0 + b_3 \cdot i
        \end{array}
      \right)
    \right)
  \\
  &=&
    tr \left(
      \left(
        \begin{array}{cc}
          \begin{array}{c}
            \left( a_0 + a_3 i \right) \cdot
              \left( b_0 - b_3 i \right) +
            \\
            \left( a_2 + a_1 i \right) \cdot
            \left( b_2 - b_1 i \right)
          \end{array}
          & \ldots \\
          \ldots &
          \begin{array}{c}
            \left( -a_2 + a_1 i \right) \cdot
              \left( -b_2 - b_1 i \right) +
            \\
            \left( a_0 - a_3 i \right) \cdot
            \left( b_0 + b_3 i \right)
          \end{array}
        \end{array}
      \right)
    \right)
  \\
  &=&
    \begin{array}{c}
      \left( a_0 + a_3 i \right) \cdot
      \left( b_0 - b_3 i \right) +
      \left( a_2 + a_1 i \right) \cdot
      \left( b_2 - b_1 i \right) +
      \\
      \left( -a_2 + a_1 i \right) \cdot
      \left( -b_2 - b_1 i \right) +
      \left(  a_0 - a_3 i \right) \cdot
      \left(  b_0 + b_3 i \right)
    \end{array}
  \\
  &=&
    \begin{array}{c}
      \left(
        a_0 b_0 +
        a_3 b_3 -
        a_0 b_3 i +
        b_0 a_3 i
      \right)
      +
      \left(
        a_2 b_2 +
        a_1 b_1 -
        a_2 b_1 i +
        a_1 b_2 i
      \right)
      \\
      \left(
        a_2 b_2 +
        a_1 b_1 +
        a_2 b_1 i -
        a_1 b_2 i
      \right)
      +
      \left(
        a_0 b_0 +
        a_3 b_3 +
        a_0 b_3 i -
        a_3 b_0 i
      \right)
    \end{array}
  \\
  &=& 2 \left( a_0 b_0 + a_1 b_1 + a_2 b_2 + a_3 b_3 \right)
  = 2 \cdot a \cdot b
\end{eqnarray}

This result is important because multiplication is an expensive operation
in computer calculation, relative to addition.  Traditionally,
calculating the trace distance between two $2 \times 2$
matrices $A$ and $B$ requires one to obtain the diagonal elements
of the product $AB$, which requires 4 complex number multiplications.
Since every complex number multiplication requires 4 real-number
multiplications, 16 real multiplications must be performed in total.
If the matrices are in the modified Pauli basis, on the other hand,
only four real multiplications are required.

The other advantage is that multiplying two gates requires only
16 real multiplications.  A traditional $2 \times 2$ matrix
multiplication, on the other hand, requires
8 complex number multiplications, or 32 real multiplications.

The final advantage is storage size: this new basis can be stored in half
the space that a full $2 \times 2$ matrix would require.

The advantages of this basis are outlined in this table:

\begin{center}
  \begin{tabularx}{0.75\textwidth}{ X | X | X | X | }
    \textbf{Task}             &
    \textbf{Regular Matrices} &
    \textbf{Pauli Basis}      &
    \textbf{Improvement}
    \\ \hline
    Find trace distance &
    16 real multiplies  &
    4 real multiplies   &
    4x speedup
    \\ \hline
    Multiply matrices &
    32 multiplies     &
    16 multiplies     &
    2x speedup
    \\ \hline
    Store a matrix &
    8 real numbers &
    4 real numbers &
    1/2 storage
    \\ \hline
  \end{tabularx}
\end{center}

\section{Future Work}

\subsection{Using multidimensional spatial indices for the
  bidirectional search middle structure}

The bidirectional search index only uses one parameter
to index the right matrices. For the reference matrices $M$ which I
chose, many matrices had similar Fowler distances to $M$. Thus, while
the algorithm was able to avoid iterating over some right matrices,
it still had to iterate over many matrices that were not
close to a given left matrix.  In fact, only .0003\% of the matrices
returned by the index were actual matches.

The modified Pauli basis offers an excellent way to parameterize the
right matrices in a spatial index:

\begin{enumerate}
  \item Its compact representation requires less space than a full
        matrix would.  In fact, since one can derive one component
        from any other three components, only three components are
        strictly required.  Space is not the only advantage; certain spatial
        indices, such as k-d trees, perform better with low-dimensional data.
        Hardware implementations of the algorithm also benefit from simpler
        calculation circuitry.

  \item Since the trace distance is just the dot product of a left
        matrix vector $a$ with a right matrix vector $b$, all right
        matrices $b$ that are close to some left matrix satisfy this
        equation:

        \begin{eqnarray}
          -D \leq a \cdot b \leq D
        \end{eqnarray}

        where $D$ is some constant related to the maximum trace distance
        between the two gates.  Geometrically, this means all of the
        close right matrices are between two parallel hyperplanes.
        The process of
        finding points between the hyperplanes should be straightforward
        to optimize. Many spatial indices group points into bounding volumes
        like boxes or spheres; checking to see if these volumes are between
        the parallel hyperplanes is a simple process.
\end{enumerate}

Libraries such as FLANN \cite{Muja2009} provide a wide variety of
spatial indices to use.

\subsection{Map-Reduce Parallelism}

The Fowler algorithm can be broken down into a cycle for each sequence
length. Each cycle is essentially a map-reduce job. During the map
phase, we assign one gate to each computer, and that computer will
consider all sequences of length $n$ which start with that gate. Once
all computers have finished the cycle, the reduce phase will merge
the data structures for unique matrices, as well as the discovered
gate sequences.

There are several advantages to map-reduce parallelism: Unique sequence
data structures can be shared with all the units between cycles. Thus,
all units can benefit from each unit\textquoteright{}s work in each
subsequent calculation cycle. If you keep track of the data structure
contents after the final stage, you can restart the algorithm from
this final stage. No specialized hardware (such as a FPGA) is required.
Anyone with access to Amazon\textquoteright{}s Elastic MapReduce service,
or a Hadoop cluster, can use a map-reduce algorithm.

Map-reduce parallelism will probably divide the algorithm\textquoteright{}s
run-time for a given sequence length by the number of computers involved.
Thus, if there are 25 computers (for 25 gates), then the algorithm
ought to run up to 25 times faster. However, since all of the computers
must merge their data after each cycle, the faster computers must
wait for the slower ones. Due to the complexity of the map-reduce
setup, this method was not implemented this quarter. However, Amazon
provides a map-reduce framework that should be straightforward to
use and scale, should someone decide to adapt the program.

\section{Related Work}

A variation of the MITM algorithm was independently invented by researchers at
the Institute for Quantum Computing at the University of Waterloo
\cite{Amy2012}.  This group also seeks to find quantum circuits of
optimal length implementing a given quantum gate.
There are a few key differences between their research and
the work presented here:

\begin{enumerate}
  \item Their work applies the algorithm to multiple-qubit gates,
        and does not combine it with Fowler's algorithm.
  \item They focus on finding \emph{exact} matches, rather than
        approximate ones.
\end{enumerate}

Their future work may benefit from the approximate matching technique
discussed in this paper, as well as the brief discussion of using spatial
indices and a change of basis to accelerate matching.  My research will
benefit from their more rigorous treatment of the algorithm, as well as its
extension to multiple qubits.

\section{Summary}

I considered a variety of optimizations to Fowler\textquoteright{}s
quantum compiler algorithm. Then, I implemented the \textquotedblleft{}meet
in the middle\textquotedblright{} algorithm in software, as well as a change
of basis technique, and I presented
the results here. While the algorithm certainly provides a dramatic
performance boost, it also requires a lot of memory to maintain the
middle index structure I introduced.
Future work involves using map-reduce parallelism and better spatial indices
to improve performance.

\section{Acknowledgements}

I performed most of this research independently, but received significant
guidance and assistance from the following individuals and organizations.
Without their involvement, this research project would not have happened!

\begin{enumerate}
  \item \textbf{Paul Pham} \textendash{} the UW graduate student
        who suggested the research topic for this
        project, and who provided essential quantum
        computing context and advice. I had weekly
        meetings with him, and I worked with him on his
        pulse sequence board two years ago. He is
        working on his own quantum compiler based on the
        Solovay-Kitaev Theorem.
  \item \textbf{Austin Fowler} \textendash{} a Research Fellow
        in Quantum
        Computer Science at the University of Melbourne.
        He wrote the original paper describing the
        sequence-skipping optimization, upon which my
        research is based. He also supplied the C source
        code to his algorithm, so that I could test
        my optimizations.
  \item \textbf{Aram Harrow} \textendash{} my faculty advisor,
        who came up with smart suggestions for error
        accumulation analysis and calculation
        optimization. He also indirectly proposed the
        MITM algorithm at the beginning of this research
        project.
\end{enumerate}

\bibliography{qco}

\end{document}